\documentclass[fleqn,twoside]{article}
\usepackage{espcrc2}
\usepackage{psfig}

\usepackage{graphicx}
\usepackage[figuresright]{rotating}

\newcommand{\AmS}{{\protect\the\textfont2
  A\kern-.1667em\lower.5ex\hbox{M}\kern-.125emS}}
\newcommand{\be}{\begin{equation}}
\newcommand{\ee}{\end{equation}}
\newcommand{\gev}{\rm GeV}
\newcommand{\kev}{\rm keV}
\newcommand{\vev}[1]{\langle #1 \rangle}

\title{Inelastic Dark Matter at DAMA, CDMS and Future Experiments}
\author{David R. Smith\address[MIT]{Center for Theoretical Physics, Massachusetts Institute of Technology,\\ Cambridge, MA, USA} and 
Neal Weiner\address[UW]{Department of Physics, University of Washington, \\ 
        Seattle, WA, USA}%
        \thanks{This work was partially supported by the DOE under contract DE-FGO3-96-ER40956.
}}
\begin{document}

\begin{abstract}
The DAMA annual modulation signature, interpreted as evidence for a spin-independent WIMP coupling, seems in conflict with null results from CDMS. However, in models of ``inelastic dark matter'',  the experiments are compatible. Inelastic dark matter can arise in supersymmetric theories as the real component of a sneutrino mixed with a singlet scalar. In contrast with ordinary sneutrino dark matter, such particles can satisfy all experimental constraints while giving the appropriate relic abundance. We discuss the modifications to the signal seen at DAMA, in particular noting the strong suppression of low energy events in both modulated and unmodulated components. We discuss future experiments, with emphasis on distinguishing inelastic dark matter from ordinary dark matter, and stressing the significance of experiments with heavy target nuclei, such as xenon and tungsten.
\vspace{1pc}
\end{abstract}

\maketitle

\section{Introduction}
Recently the DAMA collaboration has reported evidence from four years of study of an annual modulation signal consistent with a WIMP \cite{Bernabei:2000qi}. The data satisfy the six requirements of a WIMP signal in that it is a modulation of the single hit rate, contained entirely in the low energy bins, with the appropriate shape, phase, period, and amplitude that are expected. Studies of possible systematic errors have turned up no candidates to explain the modulation \cite{Bernabei:2000ew}.

At the same time, the CDMS collaboration has reported no evidence for a WIMP signal in their experiment \cite{Abusaidi:2000wg,Abrams:2002nb}, claiming a conflict with DAMA as interpreted as a WIMP coupled with spin-independent (SI) interactions at $99.8\%$, even given no assumptions about  the background \cite{Abrams:2002nb}. 

Given this conflict, and the absence of a clear source of systematic error, it is worthwhile to consider alternative forms of dark matter, with different interaction properties from the neutralino. The neutralino is a very appealing candidate, both because it arises in a motivated theory (supersymmetry) and because it naturally has the right relic abundance. In looking for a alternative theory, we would like to retain both of these features, while still explaining the discrepancy between CDMS and DAMA.

\subsection{Inelastic Dark Matter}
Let us consider ``inelastic dark matter'' (iDM) \cite{Smith:2001hy}. We assume the presence of two particles $\chi_1$ and $\chi_2$, such that $\delta = m_2-m_1 >0$. We further assume that $\chi_1$ constitutes the dark matter in the galaxy, and that $\chi_1$ can only scatter off of nuclei by making an inelastic transition to $\chi_2$. That is, the allowed scattering is $\chi_1 N\rightarrow \chi_2 N$. Let us emphasize that the inelasticity is {\em not due to a nuclear transition, but instead in the final state of the dark matter particle, itself.}

It is quite simple to construct models of this type, but we will address this later. For now, let us focus on the kinematical changes which occur in a scattering experiment. The central difference from ordinary elastic scattering is the requirement on the velocity $\beta$ for a scattering to occur:
\begin{equation}
\beta^2 > \frac{2 \delta (m_N+ m_{\chi})}{m_N m_{\chi}},
\ee
where $m_N$ is the mass of the target nucleus. The important point is that the right hand side is an {\em increasing} (more severe) function for {\em lighter} target nuclei. In the halo there is a distribution of velocities. Because the DAMA target is NaI ($A_I=127$) while the CDMS target is Ge ($A_{Ge}=73$), a larger range of velocities will be visible at DAMA than at CDMS (figure \ref{fig:dist}). As a consequence, the sensitivity of DAMA relative to CDMS is enhanced when compared with the elastic case.

\begin{figure}
\psfig{file=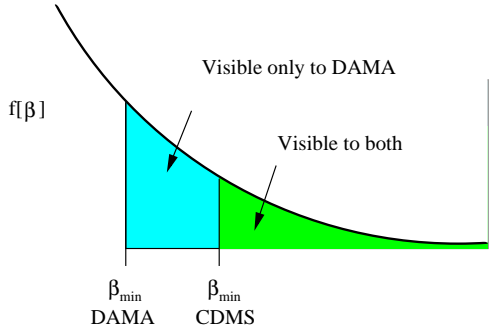,width=0.4\textwidth}
\caption{Tail of the distribution of particles in the halo as a function of velocity. Many particles are visible to DAMA but not to CDMS. \vskip -0.2in}
\label{fig:dist}
\end{figure}

A second effect enhances the sensitivity of DAMA relative to CDMS. Because we are sensitive to the number of particles above some velocity $\beta_{min}$, as we orbit the sun, the number of particles which can scatter can change considerably. As a consequence, the modulated amplitude can be considerably larger than $7\%$ of the unmodulated amplitude, as is the limit in the usual elastic case. In figure \ref{fig:modamp}, we see this enhancement.

\begin{figure}
\psfig{file=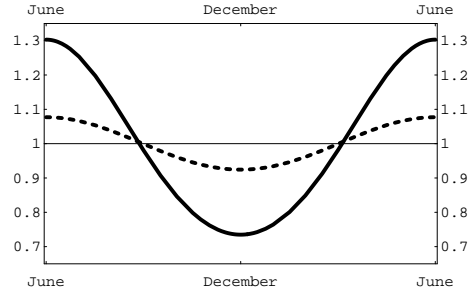,width=0.4\textwidth}
\vskip-0.2in
\caption{DAMA signal for $m_\chi=50\gev$ and elastic (dashed) and $\delta=100 \kev$  (solid). \vskip-0.2in}
\label{fig:modamp}
\end{figure}

An interesting result of this involves the interpretation of the DAMA data. It is often noted (e.g., \cite{Sadoulet:1999rq}) that the DAMA best fit point has an unmodulated signal comparable to their measured background in certain low energy bins. Here, this is no longer the case in general. In fact, the suppression of the DAMA unmodulated signal is most pronounced in the low energy bins in question, as we will discuss in the next section.

The combination of all changes allows us to fit the ``model-independent'' data of \cite{Bernabei:2000qi}, and compare with the limits of CDMS in our framework, requiring fewer than six of the observed nuclear scatterers to be WIMP events. Making certain reasonable assumptions that not too much of the DAMA signal lies in the high energy bins, we achieve the allowed parameter space of figure \ref{fig:paramspace}. We show the allowed region for $m_\chi=100 \gev$.

Recently, the DAMA collaboration has performed a complete analysis of their data within the iDM framework \cite{Bernabei:2002pp}, and found regions in good agreement with the estimated regions of \cite{Smith:2001hy}. Interestingly, the best fit points lie well in the inelastic regime, where both the spectrum and strength of modulation can differ signficiantly from the elastic case.
\vskip0.2in

\subsection{Experimental Signals of iDM}
Although inelastic dark matter offers an attractive resolution to the conflict between CDMS and DAMA, it is essential that the scenario offer additional experimental signals to distinguish it from elastically scattering dark matter.

Of great importance are the upcoming experiments using targets at least as heavy as iodine. Xenon ($A_{Xe}=131$) is being used in a number of upcoming experiments. Still, while xenon and iodine are similar in mass, the sensitivity of most experiments compared with DAMA changes as a function of $\delta$. This would be at most a factor of a few, and so a test should be available in the near future. 
Also important is the upcoming CRESST experiment, with a tungsten target ($A_W=183$) \cite{CRESST}, which would be significantly more sensitive than any of the existing experiments.
\begin{figure}
\psfig{file=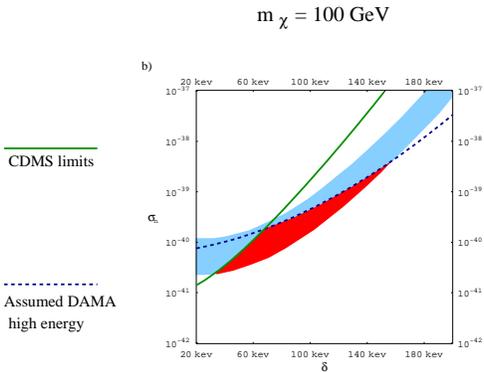,width=0.4\textwidth}
\vskip-0.2in
\caption{Allowed parameter space (dark shaded) as a function of cross section per effective neutron (see ref \cite{Smith:2001hy}) and $m_\chi$.\vskip-0.3in}
\label{fig:paramspace}
\end{figure}

The suppression of event rate is not the only significant difference. Because the kinematics are changed, the shape of the signal as a function of energy can be considerably different. In figure \ref{fig:modspec} we see how the modulated signal can appear as a function of energy at DAMA. While for certain values of $\delta$ the elastic case and inelastic case appear comparable, there can also be significant spectral differences. Currently the DAMA spectrum is unpublished, only fits to it are available. It is possible that presently existing data could help to suggest whether this scenario is correct.

At experiments which study the total rate (as opposed to modulation), the effect is even more pronounced. The signal spectrum, which for elastic WIMPs would rise dramatically at low energies, would now dip at low energies. The position of the peak would depend sensitively on $\delta$. We show an example spectrum as would be seen at CDMS in figure \ref{fig:cdm}. Such a signal would be a smoking gun of the iDM scenario. Measuring it would give great information into both the inelastic WIMP and the distribution of dark matter in the halo.
\vskip-0.2in
\section{Models of iDM}

So far we have considered iDM only as an attractive idea, but have not yet discussed how such a model would arise. In fact, as described in \cite{Smith:2001hy}, it is quite simple to realize such a model in supersymmetric theories. 

\begin{figure}
\psfig{file=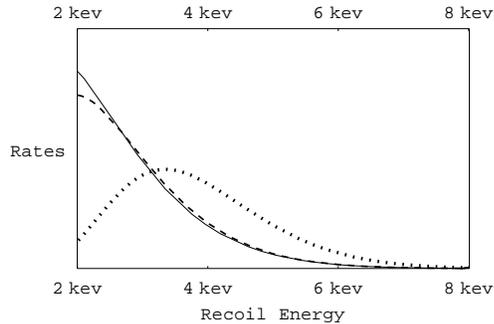,width=0.4\textwidth}
\vskip-0.2in
\caption{DAMA signal for $m_\chi=60 \gev$ and $\delta=0,100,150 \kev$.}
\vskip -0.3in
\label{fig:modspec}
\end{figure}

A complex scalar field is composed of two real scalar fields. If the scalar is charged under some gauge symmetry, one finds that the coupling of the gauge boson is {\em off-diagonal} \cite{Hall:1997ah}. That is, if $\phi=a+i b$, we have in the Lagrangian
\be
{\mathcal{L}} \supset A_\mu (a \partial^\mu b - b \partial^\mu a).
\ee
If the gauge symmetry is broken, then $a$ and $b$ will have no symmetry requiring their degeneracy, so higher order effects will naturally split their masses. This is then a general setup for achieving inelastic dark matter.

A particular example of this is a ``mixed sneutrino'', or a sneutrino which is then mixed with a singlet scalar field. Examples of models which are consistent with iDM have been given in \cite{Arkani-Hamed:2000bq,Arkani-Hamed:2000kj}. Let us consider a supersymmetry breaking spurion $X$ which acquires $A$ and $F$ component vevs of roughly the same scale
\be
\vev{X} \approx 10^{11}\gev + \theta^2 (10^{11} \gev)^2,
\ee
and consider a singlet field $N$ which acquires a mass through the usual Giudice-Masiero mechanism \cite{Giudice:1988yz} with a Lagrangian
\be
\left[ \frac{X}{M_{Pl}} LNH\right]_F +\left[\frac{X^\dagger}{M_{Pl}} NN+\frac{X^\dagger X X^\dagger}{M_{Pl}^3} N N \right]_D.
\ee
These terms lead to a scalar potential 
\be
V = A \tilde l \tilde n h + m_n^2  \tilde n \tilde n^* + \Delta^2 (\tilde n \tilde n +\tilde n^* \tilde n^*) + m_l^2 \tilde l \tilde l^*.
\ee
The lightest eigenstate is the real component of a linear combination of $\tilde l$ and $\tilde n$.  The LSP then is an inelastic WIMP with splitting $\delta \sim \Delta^2/m_{LSP}$, which is the correct size to explain the DAMA/CDMS discrepancy.

It is important to stress that the parameter $\Delta$ is a symmetry breaking parameter, without which the real components of the LSP would be degenerate. Because of this, even while $m_l^2$, $m_n^2$ and $A$ may all be renormalized, the small inelasticity parameter is stable, making the model natural. 

\begin{figure}
\psfig{file=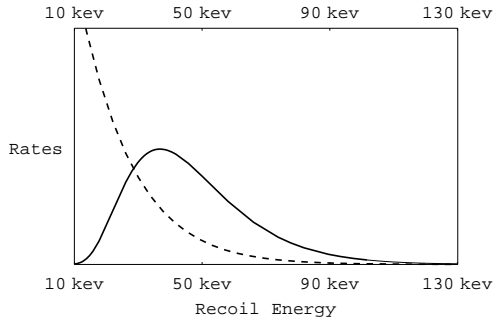,width=0.4\textwidth}
\vskip-0.2in
\caption{Spectrum at CDMS for $m_\chi=50 \gev$ and $\delta=100 \kev$.}
\vskip-0.2in
\label{fig:cdm}
\end{figure}

Because the LSP is a combination of sneutrino and the singlet scalar, its coupling to $Z$ is suppressed compared with the minimal sneutrino of the MSSM. As a consequence, it can evade direct limits from LEP and achieve the proper relic abundance \cite{Arkani-Hamed:2000bq,Smith:2001hy}.

\section{Conclusions and Outlook}
In summary, we have seen that inelastic dark matter can easily explain the discrepancy between CDMS and DAMA. Like the neutralino, inelastic dark matter can arise naturally in supersymmetric theories as the real component of a mixed sneutrino.  This well-motivated candidate is natural both in that it arises simply out of supergravity theories, and that its important features (the small splitting $\delta$) are radiatively stable.

Such models make exciting predictions for the upcoming round of dark-matter experiments. For small to moderate $\delta$, CDMS would be expected to see a signal after its move to the Sudan mine with low energy events suppressed. For moderate to large $\delta$, DAMA would see a spectral deformation from the elastic case. In all events, upcoming xenon and tungsten experiments would see a signal with varying degrees of spectral distortion.

\begin{figure}
\psfig{file=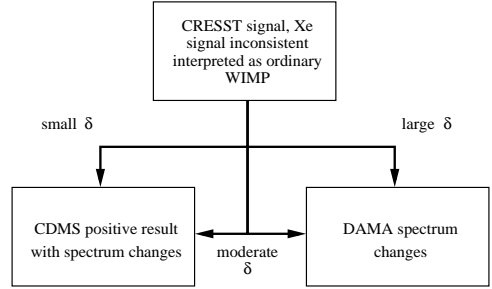,width=0.4\textwidth}
\vskip-0.25in
\caption{Possible future detection scenarios.}
\vskip-0.21in
\label{fig:scenario}
\end{figure}


\begin{thebibliography}{9}
\vskip -0.1in
\bibitem{Bernabei:2000qi}
R.~Bernabei {\it et al.}  [DAMA Collaboration],
Phys.\ Lett.\ B {\bf 480}, 23 (2000).


\bibitem{Bernabei:2000ew}
R.~Bernabei {\it et al.},
Eur.\ Phys.\ J.\ C {\bf 18}, 283 (2000).


\bibitem{Abusaidi:2000wg}
R.~Abusaidi {\it et al.}  [CDMS Collaboration],
Nucl.\ Instrum.\ Meth.\ A {\bf 444}, 345 (2000)
[Phys.\ Rev.\ Lett.\  {\bf 84}, 5699 (2000)]
[arXiv:astro-ph/0002471].


\bibitem{Abrams:2002nb}
D.~Abrams {\it et al.}  [CDMS Collaboration],
arXiv:astro-ph/0203500.


\bibitem{Smith:2001hy}
D.~R.~Smith and N.~Weiner,
Phys.\ Rev.\ D {\bf 64}, 043502 (2001)
[arXiv:hep-ph/0101138].


\bibitem{Sadoulet:1999rq}
B.~Sadoulet,
Int.\ J.\ Mod.\ Phys.\ A {\bf 15S1}, 687 (2000)
[eConf {\bf C990809}, 687 (2000)].


\bibitem{Bernabei:2002pp}
R.~Bernabei {\it et al.},
Eur.\ Phys.\ J.\ C {\bf 23}, 61 (2002).


\bibitem{CRESST}
J.~Jochum, talk given at Dark Matter 2002: Sources and Detection of Dark Matter and Dark Energy in the Universe, Marina del Rey, CA.


\bibitem{Hall:1997ah}
L.~J.~Hall, T.~Moroi and H.~Murayama,
Phys.\ Lett.\ B {\bf 424}, 305 (1998)
[arXiv:hep-ph/9712515].

\bibitem{Arkani-Hamed:2000bq}
N.~Arkani-Hamed, L.~J.~Hall, H.~Murayama, D.~R.~Smith and N.~Weiner,
Phys.\ Rev.\ D {\bf 64}, 115011 (2001)
[arXiv:hep-ph/0006312].


\bibitem{Arkani-Hamed:2000kj}
N.~Arkani-Hamed, L.~J.~Hall, H.~Murayama, D.~R.~Smith and N.~Weiner,
arXiv:hep-ph/0007001.


\bibitem{Giudice:1988yz}
G.~F.~Giudice and A.~Masiero,
Phys.\ Lett.\ B {\bf 206}, 480 (1988).



\end{thebibliography}
\end{document}